\documentclass[journal]{IEEEtran}
\ifCLASSINFOpdf
\else
\fi
\usepackage{graphicx}
\usepackage{subfig}

\usepackage{authblk}

\begin{document}
%
\title{B-FERL: Blockchain based Framework for Securing Smart Vehicles}
\author[1, 2]{Chuka Oham}
\author[1]{Regio Michellin}
\author[1]{Salil S. Kanhere}
\author[3, 2]{Raja Jurdak}
\author[1]{Sanjay Jha}
\affil[1]{University of New South Wales, Sydney, Australia \\
Email: fisrtname.lastname@unsw.edu.au
}
\affil[2]{CSIRO Data61}
\affil[3]{Queensland University of Technology, Brisbane, Australia, Email: r.jurdak@qut.edu.au}
\maketitle

\begin{abstract}
The ubiquity of connecting technologies in smart vehicles and the incremental automation of its functionalities promise significant benefits, including a significant decline in congestion and road fatalities. However, increasing automation and connectedness broadens the attack surface and heightens the likelihood of a malicious entity successfully executing an attack. In this paper, we propose a Blockchain based Framework for sEcuring smaRt vehicLes (B-FERL). B-FERL uses permissioned blockchain technology to tailor information access to restricted entities in the connected vehicle ecosystem. It also uses a challenge-response data exchange between the vehicles and roadside units to monitor the internal state of the vehicle to identify cases of in-vehicle network compromise. In order to enable authentic and valid communication in the vehicular network, only vehicles with a verifiable record in the blockchain can exchange messages. Through qualitative arguments, we show that B-FERL is resilient to identified attacks. Also, quantitative evaluations in an emulated scenario show that B-FERL ensures a suitable response time and required storage size compatible with realistic scenarios. Finally, we demonstrate how B-FERL achieves various important functions relevant to the automotive ecosystem such as trust management, vehicular forensics and secure vehicular networks.
\end{abstract}

\begin{IEEEkeywords}
\textit{CAVs}, ECUs, Blockchain, Merkle root, Forensics, Security, Vehicular network, Sensors, Trust management.
\end{IEEEkeywords}

%
\IEEEpeerreviewmaketitle

\section{Introduction}
%
%
%
%
With technological advancements in the automotive industry in recent times, modern vehicles are no longer made up of only mechanical devices but are also an assemblage of complex electronic devices called electronic control units (ECUs) which provide advanced vehicle functionality and facilitate independent decision making. ECUs receive input from sensors and runs computations for their required tasks~\cite{Alam:2018}. These vehicles are also fitted with an increasing number of sensing and communication technologies to facilitate driving decisions and to be \textit{self aware}~\cite{Anupam:2018}. However, the proliferation of these technologies have been found to facilitate the remote exploitation of the vehicle [7]. Malicious entities could inject malware in ECUs to compromise the internal network of the vehicle~\cite{Anupam:2018}. The internal network of a vehicle refers to the communications between the multiple ECUs in the vehicle over on-board buses such as the controller area network (CAN)~\cite{Han:2014}. The authors in [7] and [8] demonstrated the possibility of such remote exploitation on a connected and autonomous vehicle (CAV), which allowed the malicious entity to gain full control of the driving system and bring the vehicle to a halt.\\
To comprehend the extent to which smart vehicles are vulnerable, we conducted a risk analysis for connected vehicles in [1] and identified likely threats and their sources. Furthermore, using the Threat Vulnerability Risk Assessment (TVRA) methodology, we classified identified threats based on their impact on the vehicles and found that compromising one or more of the myriad of ECUs installed in the vehicles poses a considerable threat to the security of smart vehicles and the vehicular network. Vehicular network here refers to communication between smart vehicles and roadside units (RSUs) which are installed and managed by the transport authority. These entities  exchange routine and safety messages according to the IEEE802.11p standard [4]. By compromising ECUs fitted in a vehicle, a malicious entity could for example, broadcast false information in the network to affect the driving decisions of other vehicles. Therefore, in this paper, we focus on monitoring the state of the in-vehicle network to enable the detection of an ECU compromise. 


Previous efforts that focus on the security of in-vehicle networks have focused on intrusion and anomaly detection which enables the detection of unauthorized access to in-vehicle network [9-11], [15], [23] and the identification of deviation from acceptable vehicle behavior~\cite{Wasicek:2014}. Several challenges however persist. First, proposed security solutions are based on a centralized design which relies on a Master ECU that is responsible for ensuring valid communications between in-vehicle ECUs [9-10] [23]. However, these solutions are vulnerable to a single point of failure attack where an attacker's aim is to compromise the centralized security design. Furthermore, if the Master ECU is either compromised or faulty, the attacker could easily execute actions that undermine the security of the in-vehicle network. In-addition, efforts that focus on intrusion detection by comparing ECU firmware versions [10] [11] [15] are also vulnerable to a single point of exploitation whereby the previous version which is centrally stored could be altered. These works [11] [15] also rely on the vehicle manufacturer to ultimately verify the state of ECUs. However, vehicle manufacturers could be motivated to execute malicious actions for their benefits such as to evade liability [3].
Therefore, decentralization of the ECU state verification among entities in the vehicular ecosystem is desirable for the security of smart vehicles. Finally, the solution proposed in [24] which focuses on observing deviations from acceptable behavior utilized data generated from a subset of ECUs. However, this present a data reliability challenge when an ECU not included in the ECU subset is compromised. \\

We argue in this paper that Blockchain (BC) [12] technology has the potential to address the aforementioned challenges including centralization, availability and data reliability. \\
\textbf{BC } is an immutable and distributed ledger technology that provides verifiable record of transactions in the form of an interconnected series of data blocks. BC can be public or permissioned [3] to differentiate user capabilities including who has the right to participate in the BC network. BC replaces centralization with a trustless consensus which when applied to our context can ensure that no single entity can assume full control of verifying the state of ECUs in a smart vehicle. The decentralized consensus  provided by BC is well-suited for securing the internal network of smart vehicles by keeping track of historical operations executed on the vehicle's ECUs such as firmware updates, thus easily identifying any change to the ECU and who was responsible for that change. Also, the distributed structure of BC provides robustness to a single point of failure. 

\subsection{Contributions and Paper Layout}
Having identified the limitations of existing works, we propose a Blockchain based Framework for sEcuring smaRt vehicLes (B-FERL). B-FERL is an apposite countermeasure for in-vehicle network security that exposes threats in smart vehicles by ascertaining the state of the vehicle’s internal controls. Also, given that data modification depicts a successful attempt to alter the state of an ECU, B-FERL also suffices as a data reliability solution that ensures that a vehicle's data is trustworthy. We utilize a permissioned BC to allow only trusted entities manage the record of vehicles in the BC network. This means that state changes of an ECU are summarized, stored and managed distributedly in the BC.\\

\textit{The key contributions of this paper are summarized as follows:} \\

\textbf{(1)} We present B-FERL; a decentralized security framework for in-vehicle networks. B-FERL ascertains the integrity of in-vehicle ECUs and highlights the existence of threats in a smart vehicle. To achieve this, we define a two-tier blockchain-based architecture, which introduces an initialization operation used to create record vehicles for authentication purposes and a  challenge-response mechanism where the integrity of a vehicle's internal network is queried when it connects to an RSU to ensure its security.\\

\textbf{(2)} We conduct a qualitative evaluation of B-FERL to evaluate its resilience to identified attacks. We also conduct a comparative evaluation with existing approaches and highlight the practical benefits of B-FERL.  Finally, we characterize the performance of B-FERL via extensive simulations using the CORE simulator against key performance measures such as the time and storage overheads for smart vehicles and RSUs.\\

\textbf{(3)} Our proposal is tailored to meet the integrity requirement for securing smart vehicles and the availability requirement for securing vehicular networks and we provide succinct discussion on the applicability of our proposal to achieve various critical automotive functions such as vehicular forensics, secure vehicular communication and trust management. \\

This paper is an extension of our preliminary ideas presented in [1]. Here, we present a security framework for detecting when an in-vehicle network compromise occurs and provide evidence that reflect actions on ECUs in a vehicle. Also, we present extensive evaluations to demonstrate the efficacy of B-FERL. \\
The rest of the paper is structured as follows. In section 2, we discuss related works and present an overview of our proposed framework in Section 3 where we describe our system, network and threat model. Section 4 describes the details of our proposed framework. In section 5, we discuss results of the performance evaluation. Section 6 present discussions on the potential use cases of B-FERL,  comparative evaluation with closely related works, and we conclude the paper in Section 7. 

\section{Related Work}
BC has been proposed as security solutions for vehicular networks. However, proposed solutions have not focused on the identification of compromised ECUs for securing vehicular networks.
The author in~\cite{Blackchain:2017} proposed Blackchain, a BC based message revocation and accountability system for secure vehicular communication. However, their proposal does not consider the reliability of data communicated in the vehicular network which could be threatened when an in-vehicle ECU is compromised. The author in~\cite{Ali:2017} presents a BC based architecture for securing automotive networks. However they have not described how their architecture is secured from insider attacks where authorised entities could be motivated to execute rogue actions for their benefits. Also, their proposal does not consider the veracity of data from vehicles. The authors in~\cite{cube:2018} proposed a security platform for autonomous vehicle based on blockchain but have not presented a description of their architecture and its applicability for practical scenarios. Also, their security is towards the prevention of unauthorized network entry using a centralized intrusion detector which is vulnerable to a single point of failure attack. Their proposal do not also consider the malicious tendencies of authorized entities as described in~\cite{Oham:2018}. 

The authors in~\cite{Coin:2018} proposed CreditCoin; a privacy preserving BlockChain based incentive announcement and reputation management scheme for smart vehicles. Their proposal is based on threshold authentication where a number of vehicles agree on a message generated by a vehicle and then the agreed message is sent to a nearby roadside unit. However, in addition to the possibility of collusion attacks, the requirement that vehicles would manage a copy of the blockchain presents a significant storage and scalability constraint for vehicles. The authors in~\cite{BARS:2018} have proposed a Blockchain-based Anonymous Reputation System (BARS) for Trust Management in VANETs however, they have not presented details on how reputation is built for vehicles and have also not presented justifications for their choice of reputation evaluation parameters. The authors in~\cite{Contract:2018} have proposed an enhanced Delegated Proof-of-stake (DPoS) consensus scheme with a two-stage soft security solution for secure vehicular communications. However, their proposal is directed at establishing reputation for road side infrastructures and preventing collusion attacks in the network. These authors~\cite{Coin:2018}~\cite{BARS:2018}~\cite{Contract:2018} have also not considered the security of in-vehicle networks. 

\section{B-FERL Overview and Threat Model}
In this section, we present a brief overview of B-FERL including the roles of interacting entities, and a description of the network and threat models. 

\subsection{Architecture overview}
\begin{figure*}[h]
\centering
\includegraphics[width=0.9\textwidth]{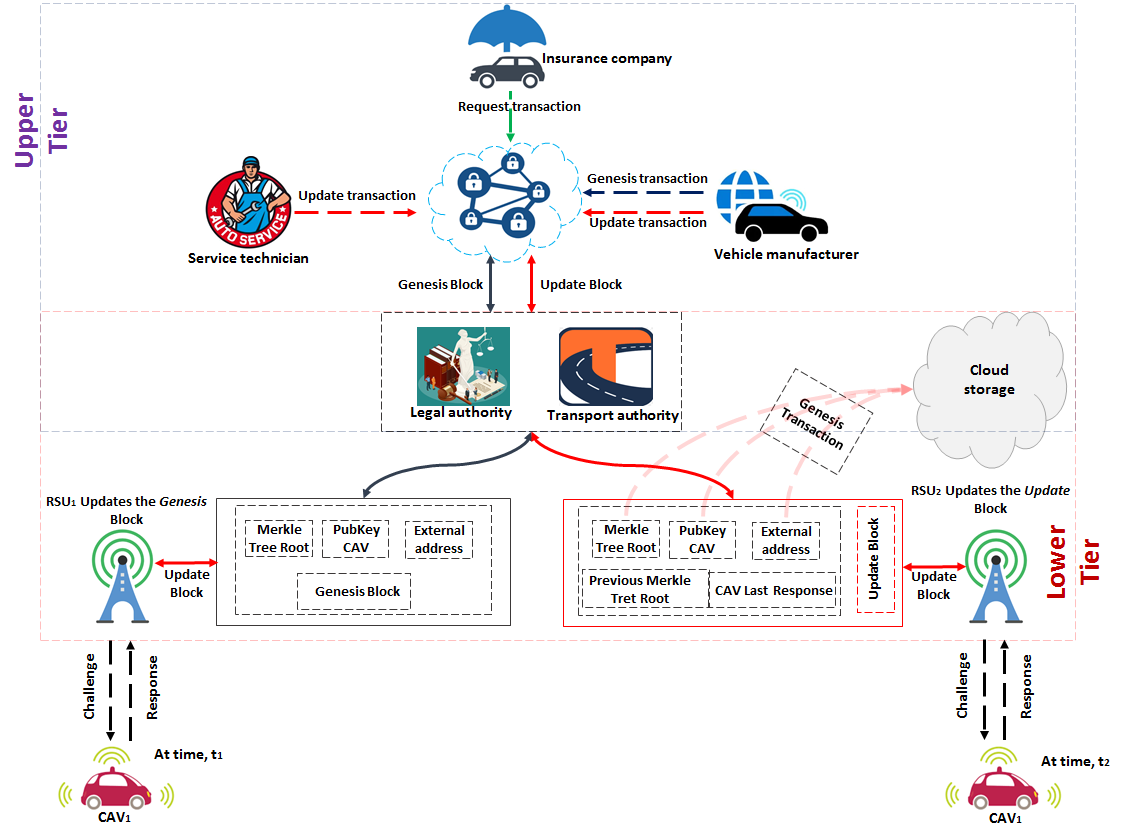}
\caption{The Proposed Blockchain Framework}
\label{fig:framework}
\end{figure*}

The architecture of our proposed security solution (B-FERL) is described in Figure~\ref{fig:framework}. 
Due to the need to keep track of changes to ECU states and to monitor the behaviour of a vehicle while operational, B-FERL consists of two main BC tiers namely, upper and lower tiers. Furthermore, these tiers clarify the role of interacting entities and ensure that entities are privy to only information they need to know. 

The upper tier comprises vehicle manufacturers, service technicians, insurance companies, legal and road transport authorities. The integration of these entities in the upper tier makes it easy to also keep track of actions executed by vehicle manufacturers and service technicians on ECUs such as firmware updates which changes the state of an ECU and allows only trusted entities such as transport and legal authorities to verify such ECU state changes.  Interactions between entities in this tier focus on vehicle registration and maintenance. The initial registration data of a vehicle is used to create a record (block) for the vehicle in the upper tier. This record stores the state of the vehicle and the hash values of all ECUs in the vehicle and is used to perform vehicle validation in the lower tier BC. This is accomplished by comparing the current state of the vehicle and the firmware hashes of each ECU in the vehicle to their values in the lower tier BC. Also, the upper tier stores scheduled maintenance or diagnostics data that reflects the actions of vehicle manufacturers and service technicians on a smart vehicle. This information is useful for the monitoring of the vehicle while operational and for making liability decisions in the multi-entity liability attribution model~\cite{Oham:2018}.\\
In the following, we describe actions that trigger interactions in the upper tier. In the rest of the paper unless specifically mentioned, we refer to smart vehicles as \textit{CAVs}.

\begin{itemize}
    \item When a \textit{CAV} is assembled, the vehicle manufacturer obtains the ECU Merkle root value ($SS_{ID}$) by computing hash values of all ECUs in the vehicle and forwards this value to the road transport and legal authorities to create a public record (block) for the vehicle. This record is utilized by RSUs to validate vehicles in the lower tier. We present a detailed description of this process in Section 3. 
    
   \item When a maintenance occurs in the vehicle, vehicle manufacturers or service technicians follow the process of obtaining the updated $SS_{ID}$ value above and communicate this to the transport and legal authorities to update the record of the vehicle and assess the integrity of its ECUs. We present a detailed description of this process in Section 3. Maintenance here means any activity that alters the state of any of the vehicle's ECUs. 
\end{itemize}

The lower tier comprises roadside units (\textit{RSUs}), smart vehicles, legal and road transport authorities. Interactions in this tier focus on identifying when an ECU in a vehicle has been compromised. To achieve this, a vehicle needs to prove its ECUs firmware integrity whenever it connects to an \textit{RSU}. When a vehicle approaches the area of coverage of an \textit{RSU}, the \textit{RSU} sends the vehicle a challenge request to prove the state of its ECUs. To provide a response, the vehicle computes the cumulative hash value of all of its ECUs i.e. its ECU Merkle root ($SS_{ID}$). The response provided by the vehicle is then used to validate its ECUs current state in comparison to the previous state in the lower tier. Also, as a vehicle moves from one \textit{RSU} to the other, an additional layer of verification is added by comparing the time stamps of its current response to the previous response to prevent the possibility of a replay attack. It is noteworthy, that compared to traditional BC which executes a consensus algorithm in order to insert transactions into a block, B-FERL relies on the appendable block concept (ABC) proposed in~\cite{Michelin:2018} where transactions are added to the blocks by valid block owners represented by their public key. Therefore, no consensus algorithm is required in B-FERL to append transactions to the block. To ensure that the integrity of a block is not compromised, ABC decouples the block header from the transactions to enable network nodes store transactions off-chain without compromising block integrity. Furthermore, to ensure scalability in the lower tier, we only store two transactions (which represents the previous and current ECU's firmware state) per vehicle and push other transactions to the cloud where historical data of the vehicle can be accessed when necessary. 
However, this operation could introduce additional latency for pushing the extra transaction from the RSU to the cloud storage. This further imposes an additional computing and bandwidth requirement for the RSU.  \\
Next, we discuss our network model which describes interacting entities in our proposed framework and their roles. 
\subsection{Network model}
To restrict the flow of information to only concerned and authorized entities, we consider a two-tiered network model as shown in Figure \ref{fig:framework}. The upper tier features the road transport and legal authorities responsible for managing the vehicular network. This tier also integrates entities responsible for the maintenance of vehicles such as vehicle manufacturers and the service technicians. It could also include auto-insurance companies who could request complimentary evidence from Transport and Legal authorities for facilitating liability decisions. For simplicity we focus on single entities for each of these however, our proposal is generalizable to the case when there are several of each entity.\\
The lower tier features \textit{CAVs} as well as RSUs which are installed by the road transport authority for the management and monitoring of traffic situation in the road network. 
For interactions between \textit{CAVs} and RSUs , we utilize the IEEE802.11p communication standard which has been widely used to enable vehicle-to-vehicle and vehicle-to-infrastructure communications [4]. However, 5G is envisaged to bring about a new vehicular communication era with higher reliability, expedited data transmissions and reduced delay [5]. Also, we utilise PKI to issue identifiable digital identities to entities and establish secure communication channels for permissible communication. 
The upper tier features a permissioned blockchain platform managed by the road transport and legal authorities. Vehicle manufacturers and service technicians participate in this BC network by sending sensor update notification transactions which are verified and validated by the BC network managers. Insurance companies on the other hand participate by sending request transactions for complimentary evidence to facilitate liability attribution and compensation payments. The lower tier also features a permissioned BC platform  managed by the road transport, legal authorities and RSUs. In this tier, we maintain vehicle-specific profiles. To achieve this, once a vehicle enters the area of coverage of a roadside unit (RSU), the RSU sends a challenge request to the vehicle by which it reports the current state of its ECUs. Once a valid response is provided, the vehicle is considered trustworthy until another challenge-response activity. \\
We present a full description of the entire process involved in our proposed framework in section 3. 
\subsection{Threat Model}
Given the exposure of \textit{CAVs} to the Internet, they become susceptible to multiple security attacks which may impact the credibility of data communicated by a vehicle. In the attack model, we consider how relevant entities could execute actions to undermine the proposed framework. The considered attacks include: \\
 \textbf{Fake data:} A compromised vehicle could try to send misleading information in the vehicular network for its benefit. For example, it could generate false messages about a traffic incident to gain advantage on the road. Also, to avoid being liable in the case of an accident, a vehicle owner could manipulate an ECU to generate false data.\\
\textbf{Code injection:} Likely liable entities such as the vehicle manufacturer and service technician could send malware to evade liability. vehicle owners on the other hand could execute such actions to for example reduce the odometer value for the vehicle to increase its resale value.\\
\textbf{Sybil attack:} A vehicle could create multiple identities to manipulate vehicular network, for example by creating false alarm such as false traffic jam etc.\\
\textbf{Masquerade attack (fake vehicle):} A compromised roadside unit could create a fake vehicle or an external adversary could create a fake vehicle for the purpose of causing an accident or changing the facts of an accident. \\
\textbf{ECU State Reversal Attack: } A vehicle owner could extract the current firmware version of an ECU and install its malicious version and revert to the original version for verification purpose. 

\section{Blockchain based Framework for sEcuring smaRt vehicLes (B-FERL)} \label{sec:b-ferl}
This section outlines the architecture of the proposed framework. As described in Figure~\ref{fig:framework}, entities involved in our framework include vehicle manufacturers, service technicians, insurance companies, \textit{CAVs}, RSUs, road transport and legal authorities. Based on entity-roles described in section 2, we categorize entities as verifiers and proposers. Verifiers are entities that verify and validate data sent to the BC. Verifiers in B-FERL include RSUs, road transport and legal authorities. Proposers are entities sending data to the BC or providing a response to a challenge request. Proposers in our framework include \textit{CAVs}, vehicle manufacturers, service technicians and insurance companies. \\
In B-FERL architecture, we assume that the CAVs are producing many transactions, especially in high density smart city areas. Most of blockchains implementations are designed to group transactions, add them into a block and only after that append the new block into the blockchain, which leads to a sequential transaction insertion. To tackle this limitation, in B-FERL we adopted a blockchain framework presented by Michelin et al.~\cite{Michelin:2018} which introduces the appendable block concept (ABC). This blockchain solution enables multiple CAVs to append transactions in different blocks at same time. The framework identifies each CAV by its public key, and for each different public key, a block is created in the blockchain data structure. The block is divided in two distinct parts: (i) block header, which contains the CAV public key, the previous block header hash, the timestamp; (ii) block payload, where all the transactions are stored. The transaction storage follows a linked list data structure, the first transaction contains the block header hash, while the coming transactions contain the previous transaction hash. This data structure allows the solution to insert new transaction into existing blocks. Each transaction must be signed by the CAV private key, once the transaction signature is validated with the block's public key, the RSU can proceed appending the transaction into the block identified by the CAV public key. Based on the public key, the BC maps all the transactions from a specific entity to the same block.  

\subsection{Transactions}
Transactions are the basic communication primitive in BC for the exchange of information among entities in B-FERL. 
Having discussed the roles of entities in each tier in B-FERL, in this section, we discuss the details of communication in each tier facilitated by the different kind of transactions. Transactions generated are secured using cryptographic hash functions (SHA-256), digital signatures and asymmetric encryption. \\
\textbf{\textit{Upper tier}}\\
Upper tier transactions include relevant information about authorized actions executed on a \textit{CAV}. They also contain interactions that reflect the time a vehicle was assembled. Also, in this tier, insurance companies could seek complementary evidence from the road transport and legal authorities in the event of an accident hence, a request transaction is also sent in this tier.  \\
\begin{figure}[h]
\centering
\includegraphics[width=0.5\textwidth]{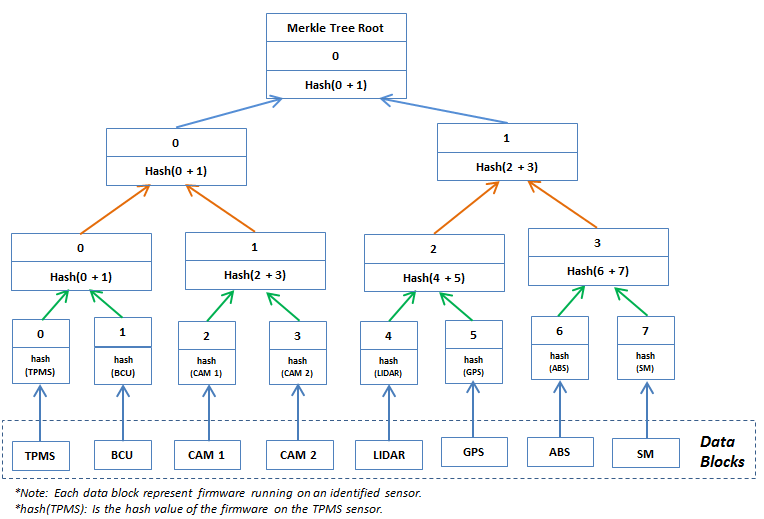}
\caption{Obtaining the Merkle tree root value}
\label{fig:merkle}
\end{figure}
\textbf{Genesis transaction:} This transaction is initiated by a vehicle manufacturer when a vehicle is assembled. The genesis transaction contains the initial $SS_{ID}$ value which is the Merkle tree root from the \textit{CAV's} ECU firmware hashes at \textit{CAV} creation time, time stamp, firmware hashes of each ECU and associated timestamps, ($H(ECU){_1}$, $T{_1}$),  ($H(ECU){_2}$, $T{_2}$), .....($H(ECU){_n}$, $T{_n}$) which reflect when an action was executed on the ECU, the public key and signature of the vehicle manufacturer. Figure \ref{fig:merkle} shows how the $SS_{ID}$ of a \textit{CAV} with 8 ECUs is derived. 
\begin{center}
    Genesis = [$SS_{ID}$, TimeStamp, ($H(ECU){_1}$, $T{_1}$),  ($H(ECU){_2}$, $T{_2}$), .....($H(ECU){_n}$, $T{_n}$), PubKey, Sign]
\end{center}
 The genesis transaction is used by the transport and legal authorities to create a genesis block for a \textit{CAV}. This block is a permanent record of the \textit{CAV} and used to validate its authenticity in the lower tier. It contains the genesis transaction, public key of the \textit{CAV}, time stamp which is the time of block creation and an external address such as an address to a cloud storage where \textit{CAV} generated data would be stored as the block size increases. \\
\textbf{Update transaction:} This transaction could be initiated by a vehicle manufacturer or a service technician. It is initiated when the firmware version of an ECU in the \textit{CAV} is updated during  scheduled maintenance or diagnostics. An update transaction leads to a change in the initial $SS_{ID}$ value and contains the updated $SS_{ID}$ value, time stamp, public key of \textit{CAV}, public key of vehicle manufacturer or service technician and their signatures. \\
When an update transaction is received in the upper tier, the update transaction updates the record (block) of the \textit{CAV} in the lower tier. The updated \textit{CAV} block will now be utilized by RSUs to validate the authenticity of the \textit{CAV} in the lower tier.\\
\textbf{Request transaction:} This transaction is initiated by an insurance company to facilitate liability decisions and compensation payments. It contains the signature of the insurance company, the data request and its public key.\\
\textbf{\textit{Lower tier}} \\
Communication in the lower tier reflect how transactions generated in the upper tier for CAVs are appended to their public record (block) in the lower tier. Additionally, we describe how the block is managed by an RSU in the lower tier and the transport and legal authorities in the upper tier.  Lower tier communications also feature the interactions between \textit{CAVs} and RSUs and describes how the integrity of ECUs in a \textit{CAV} is verified. In the following, we describe the interactions that occur in the lower tier. \\
\textbf{Updating CAV block:} Updating the block of a \textit{CAV} is either performed by the road transport and legal authorities or by an RSU. It is performed by the road transport and legal authorities after an update transaction is received in the upper tier. It is performed by an RSU after it receives a response to a challenge request sent to the vehicle. The challenge-response scenario is described in the next type of transaction. The update executed by an RSU contains a \textit{CAV’s} response which includes the signature of the \textit{CAV}, time stamp, response to the challenge and \textit{CAV’s} public key. It also contains the hash of the previous transaction in the block computed by the RSU, the signature and public key of the RSU.\\
\textbf{Challenge-Response transaction:} The Challenge-Response transaction is a request from an RSU to prove the integrity of its ECUs. This request is received  when the \textit{CAV} comes into the RSU's area of coverage. When this occurs, the \textit{CAV} receives a twofold challenge from the RSU. First is a  challenge to compute its $SS_{ID}$ to ascertain the integrity of its state. Next challenge is to compute the hash value of randomly selected ECUs to prevent and detect the malicious tendencies of vehicle owners discussed in Section 3.\\ 
The \textit{CAV} responds by providing a digitally signed response to the request.
\subsection{Operation}
In this section we describe key operations in our proposed framework. The proposed framework works in a permissioned mode where road transport and legal authorities have rights to manage the BC in the upper and lower tier. Service technicians as well as vehicle manufacturers generate data when they execute actions that alters the internal state of a \textit{CAV} while \textit{CAVs} prove the integrity of their ECUs when they connect to a RSU. \\
We define 2 critical operations in our proposed framework: 
\subsubsection{Initialization} Describes the process of creating a record for a vehicle in the vehicular network. Once a genesis transaction is generated for a \textit{CAV} by a vehicle manufacturer, upper tier verifiers verify the transaction and upon a successful verification, a genesis block is broadcasted in the lower tier for the \textit{CAV}. \\
\begin{figure}[h]
\centering
\includegraphics[width=0.53\textwidth]{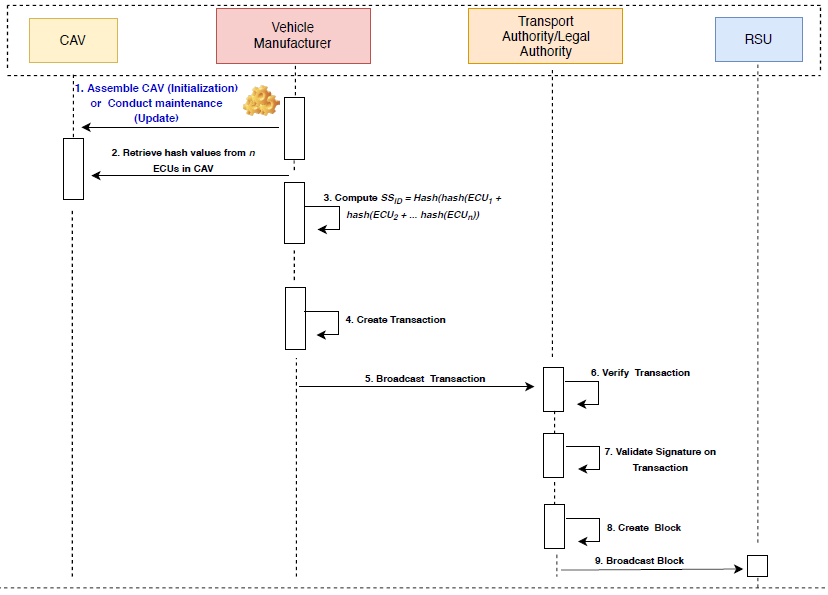}
\caption{\textit{CAV} record initialization (black) and upper-tier update (blue) operations.}
\label{fig:operation}
\end{figure}
Figure \ref{fig:operation} describes the process of block creation (assembling) for \textit{CAVs}. It outlines the requisite steps leading to the creation of a block (record) for a \textit{CAV}. 
 
\subsubsection{Update} 
\begin{figure*}[h]
\centering
\includegraphics[width=0.85\textwidth]{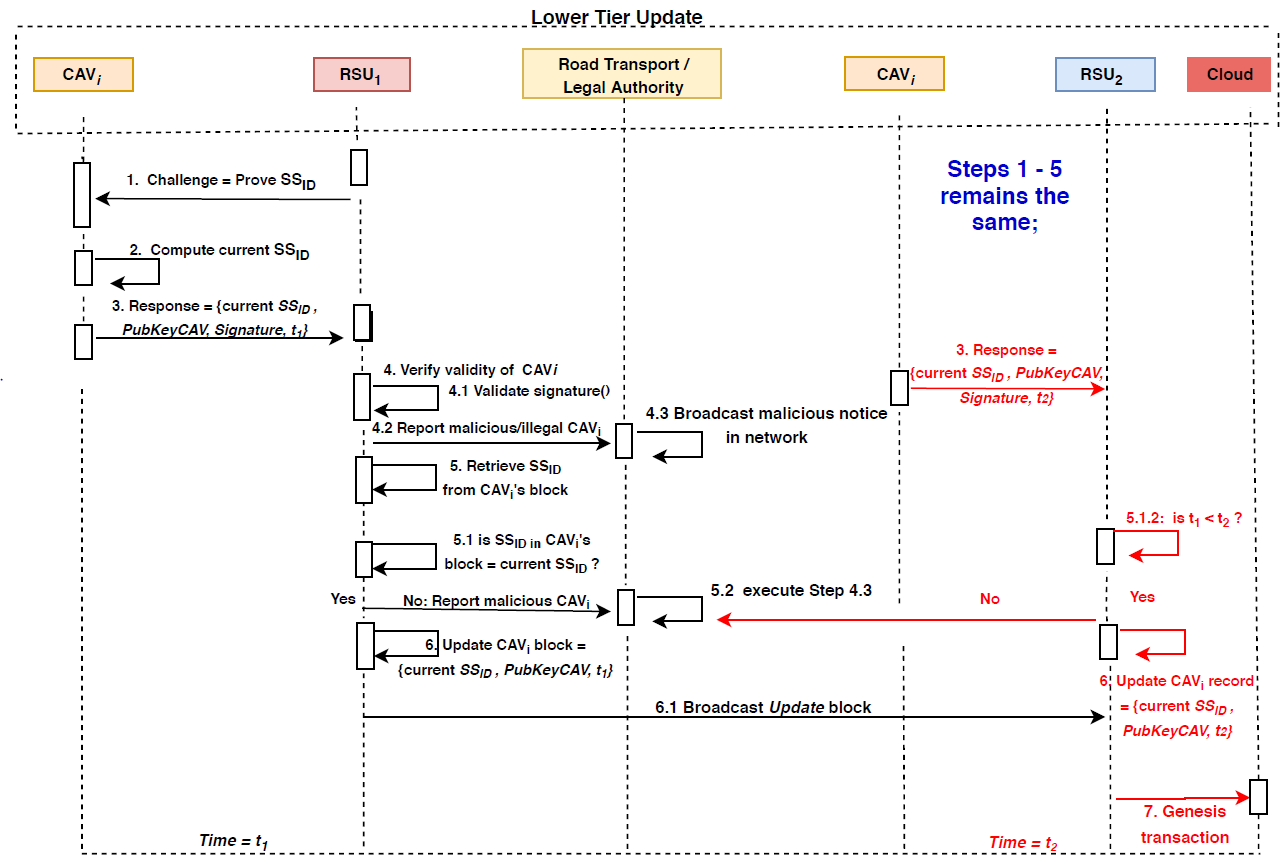}
\caption{Lower-tier update operations.}
\label{fig:lowupdate}
\end{figure*}

Describes the process of updating the record of the vehicle in the vehicular network. The update operation results in a change in the block of a \textit{CAV} in the lower tier. The update operation occurs in the upper and lower tier. In the upper tier, an update operation occurs when a vehicle manufacturer performs a diagnostic on a \textit{CAV} or when a scheduled maintenance is conducted by a service technician. In the lower tier, it occurs when a \textit{CAV} provides a response to the challenge request initiated by an RSU. In the following we discuss the update operation that occurs at both tiers. \\
\textbf{Upper-tier update:} Here, we describe how the earlier mentioned actions of the vehicle manufacturer or service technician alters the existing record for a \textit{CAV} in the vehicular network.\\
Figure \ref{fig:operation} outlines the necessary steps to update the record of a vehicle. After completing the diagnostics or scheduled maintenance (step 1), the vehicle manufacturer or service technician retrieves the hash of all sensors in the vehicle (step 2) and computes a new ECU Merkle root value (step 3). Next, an update transaction is created to reflect the action on the vehicle (step 4). This transaction includes the computed ECU Merkle root value, time stamp to reflect when the diagnostics or maintenance was conducted, signature of the entity conducting the maintenance or diagnostics and a metadata field that describes what maintenance or diagnostics was conducted on the \textit{CAV}. Next, the transaction is broadcasted in the upper tier (step 5) and verified by verifiers (step 6); road transport and legal authorities by validating the signature of the proposer (step 7). Upon signature validation, an update block is created by the verifiers for the \textit{CAV} (step 8) and broadcasted in the lower tier (step 9). \\
\textbf{Lower tier update:} We describe here how the update of a \textit{CAV’s} record is executed by an RSU after the initialization steps in the lower tier. \\
Figure \ref{fig:lowupdate} describes the necessary steps involved in updating the record of the \textit{CAV} in the lower tier. When a \textit{CAV} approaches the area of coverage of an RSU, the RSU sends the \textit{CAV} a challenge request which is to prove that it is a valid \textit{CAV} by proving its current sensor state (Step 1). For this, the \textit{CAV} computes its current $SS_{ID}$ value as well as the hash values of selected ECUs (Step 2) and forward it to the RSU including its signature, time stamp and public key (Step 3).  \\
When the RSU receives the response data from the \textit{CAV}, it first verifies that the vehicle is a valid \textit{CAV} by using its public key ($PubKey_{CAV}$) to check that the vehicle has a block in the BC (Step 4). Only valid vehicles have a block (record) in the BC. When the RSU retrieves $PubKey_{CAV}$, it validates the signature on the response data (Step 4.1). If validation succeeds, the RSU retrieves the firmware hash value in the \textit{CAV’s} block (Step 5) proceeds to compare the computed hash values with the value on the \textit{CAV’s} block (Step 5.1). Otherwise, the RSU reports to the road transport and legal authorities of the presence of a malicious \textit{CAV} or an illegal \textit{CAV} if there is no block for such \textit{CAV} in the BC (Step 4.2). If the comparison of hash values succeeds, the RSU updates the \textit{CAV’s} record in the lower tier to include the $SS_{ID}$ value, the time stamp, and public key of the \textit{CAV} (Step 6). This becomes the latest record of the \textit{CAV} in the lower tier until another challenge-response round or another maintenance or diagnostic session. However, if the hash value differs, the RSU reports to the road transport and legal authorities of the presence of a malicious \textit{CAV} (Step 5.2). \\
When the \textit{CAV} encounters another RSU, another challenge-response activity begins. This time, the RSU repeats the steps (1-5), in-addition, another layer of verification is executed. The RSU compares the time stamp on the response data to the immediate previous record stored on the lower tier blockchain (Step 5.1.2). The time stamp value is expected to continuously increase as the vehicle travels, if this is the case, RSU executes updates the \textit{CAV’s} block (Step 6). Otherwise, the RSU can detect a malicious action and report this to the road transport and legal authority (Step 5.2). However, if a malicious \textit{CAV} reports a time stamp greater than its previous time stamp, we rely on the assumption that one or more of its ECUs would have been compromised and so it would produce an $SS_{ID}$ different from its record in the lower tier. Another alternative is to comparatively evaluate its time-stamp against the time-stamp of other vehicles in the RSU area of coverage.  To ensure that the blockchain in the lower tier scales efficiently, we store only two transactions per \textit{CAV} block. In this case, after successfully executing (Step 5.1.2), the RSU removes the genesis transaction from the block and stores it in a cloud storage which can be accessed using the external address value in the \textit{CAV’s} block. \\
With the challenge-response activity, we build a behaviour profile for \textit{CAV’s} and continuously prove the trustworthiness of a vehicle while operational. Also, by keeping track of the actions of likely liable entities such as the service technician and vehicle manufacturer and by storing vehicle’s behaviour profile in the blockchain, we obtain historical proof that could be utilised as contributing evidence for facilitating liability decisions. 
\section{Performance Evaluation}
\begin{figure}[h]
\centering
\includegraphics[width=0.5\textwidth]{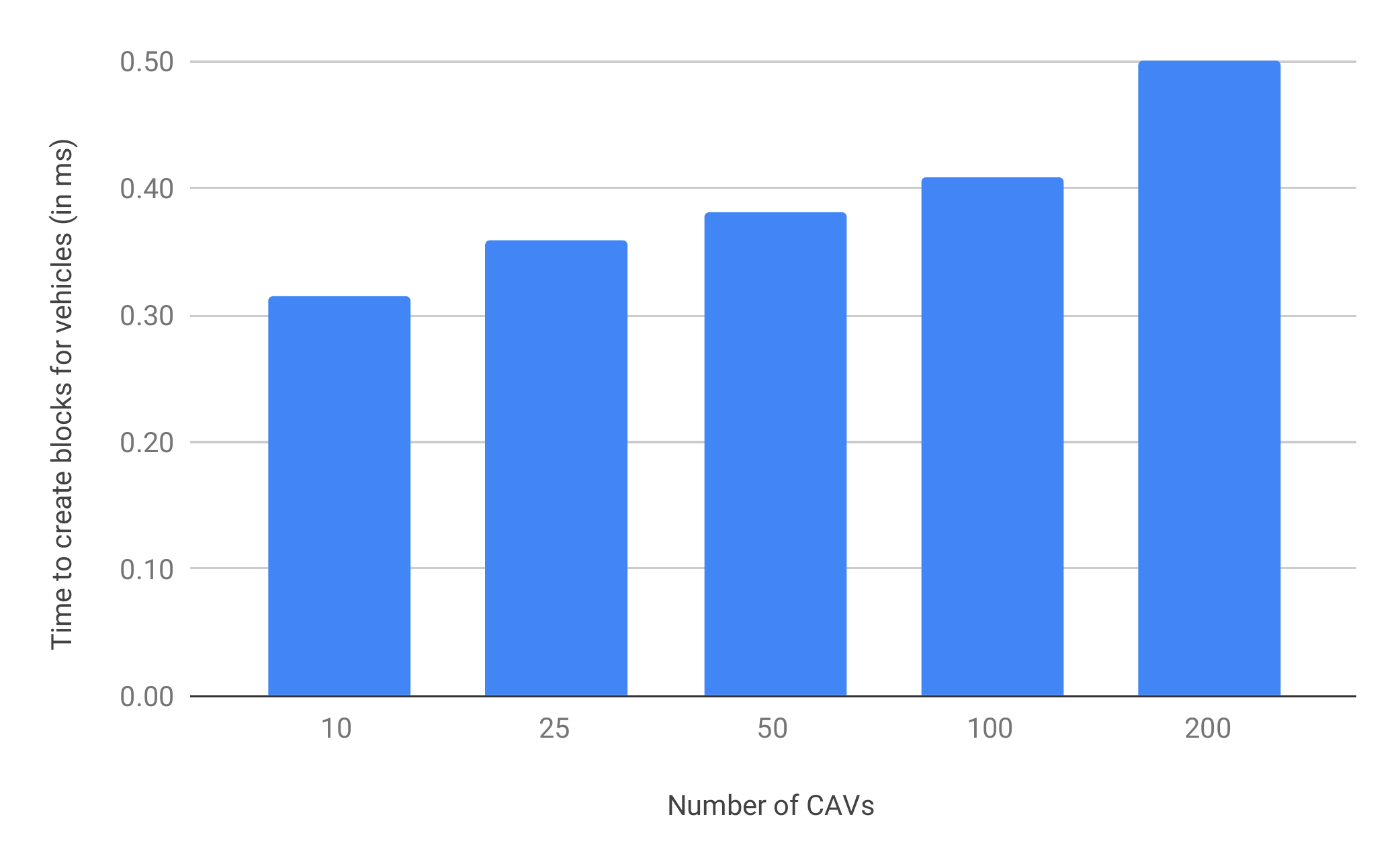}
\caption{Time taken to create a block}
\label{fig:createBlock}
\end{figure}

\begin{figure}[h]
\centering
\includegraphics[width=0.5\textwidth]{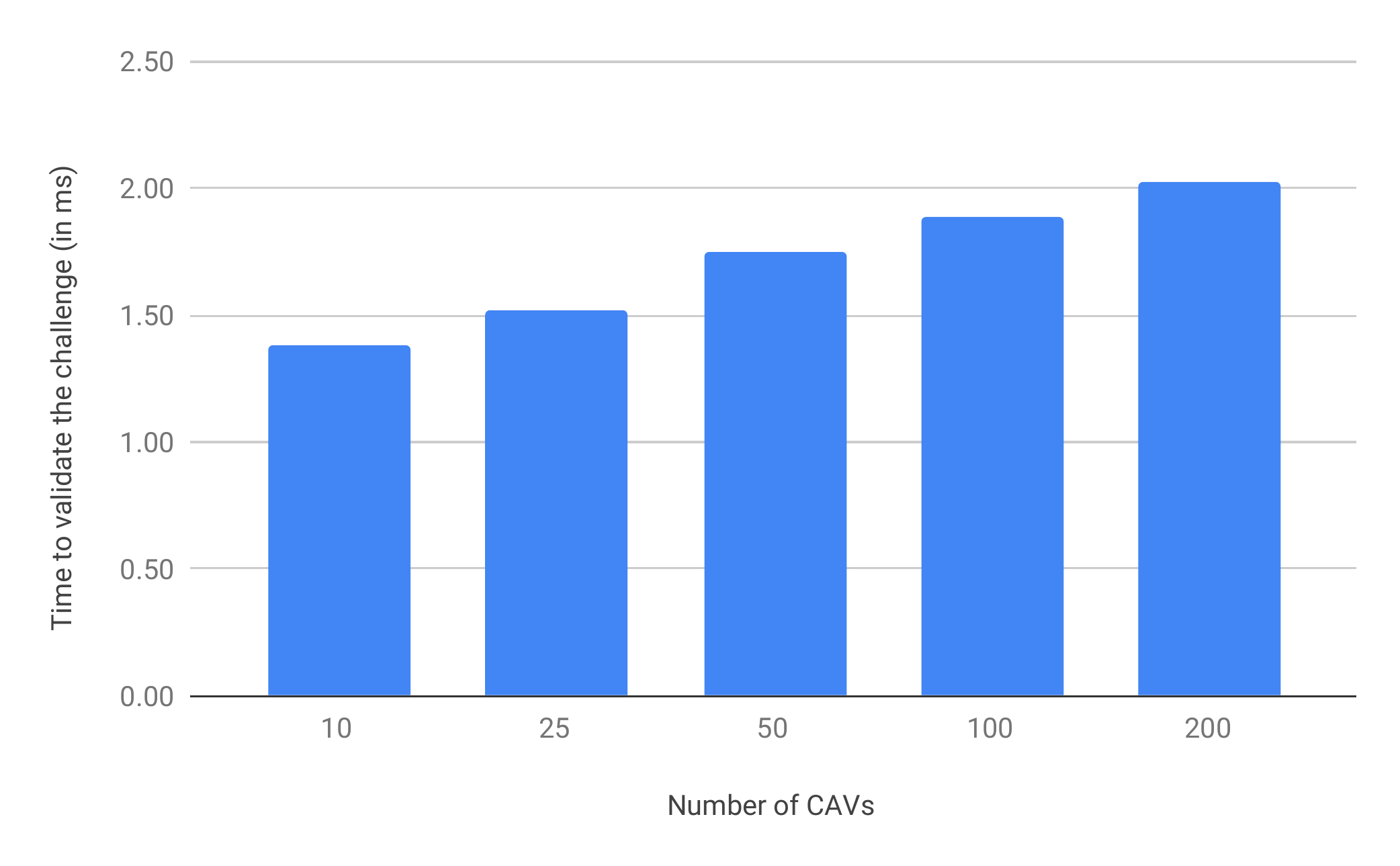}
\caption{Time taken to validate a challenge from vehicles}
\label{fig:validateChallenge}
\end{figure}

The evaluation of B-FERL was performed in an emulated scenario using Common Open Research Emulator (CORE), running in a Linux Virtual Machine using six processor cores and 12 Gb of RAM. Based on the appendable blocks concept described in section~\ref{sec:b-ferl}, B-FERL supports adding transactions of a specific \textit{CAV} to a block. This block is used to identify the \textit{CAV} in the lower tier and stores all of its records. \\
The initial experiments aim to identify the project viability, and thus enable us to plan ahead for real world scenario experimentation. The evaluated scenario consists of multiple CAVs (varying from 10 to 200) exchanging information with a peer-to-peer network with five RSU in the lower tier.
Initially, we evaluate the time it takes B-FERL to perform the system initialization. This refers to the time it takes the upper tier validators to create a record (block) for a \textit{CAV}.  Recall that in Figure~\ref{fig:operation}, creating a record for a \textit{CAV} is based on the successful verification of the genesis transaction sent from vehicle manufacturers. The results presented are the average of ten runs and we also show the standard deviation for each given scenario. In this first evaluation, we vary the amount of genesis transactions received by validators from 10 to 200 to identify how B-FERL responds to the increasing number of simultaneous transactions received. 
The results are presented in Figure~\ref{fig:createBlock}. Time increases in a linear progression as the number of \textit{CAVs} increases. The time measured in milliseconds increases from 0.31 ms (standard deviation 0.12 ms) for 10 \textit{CAVs}, to 0.49 ms (standard deviation 0.22 ms) for 200 \textit{CAVs} which is still relatively low compared to the scenario with 10 \textit{CAVs}.\\
Once the blocks are created for the \textit{CAVs}, the upper tier validators broadcast the blocks to the RSUs. In the next evaluation, we measure the time taken for an RSU to update its BC with the new block. The time required for this action is 0.06 ms for 200 \textit{CAVs} which reflects the efficiency of B-FERL given the number of \textit{CAVs}.
The next evaluation was the time that each RSU takes to evaluate the challenge response. This is an important measure in our proposed solution as it reflects the time taken by an RSU to verify the authenticity of a \textit{CAV} and conduct the ECU integrity check. This process is described in steps 4 to 6 presented in Figure~\ref{fig:lowupdate}. Figure~\ref{fig:validateChallenge} presents the average time, which increases linearly from 1.37 ms (standard deviation 0.15 ms) for 10 \textit{CAVs} to 2.02 ms (standard deviation 0.72 ms) for 200 \textit{CAVs}. From the result, we can see that the actual values are small even for large group of \textit{CAVs}.
\begin{figure}[h]
\centering
\includegraphics[width=0.5\textwidth]{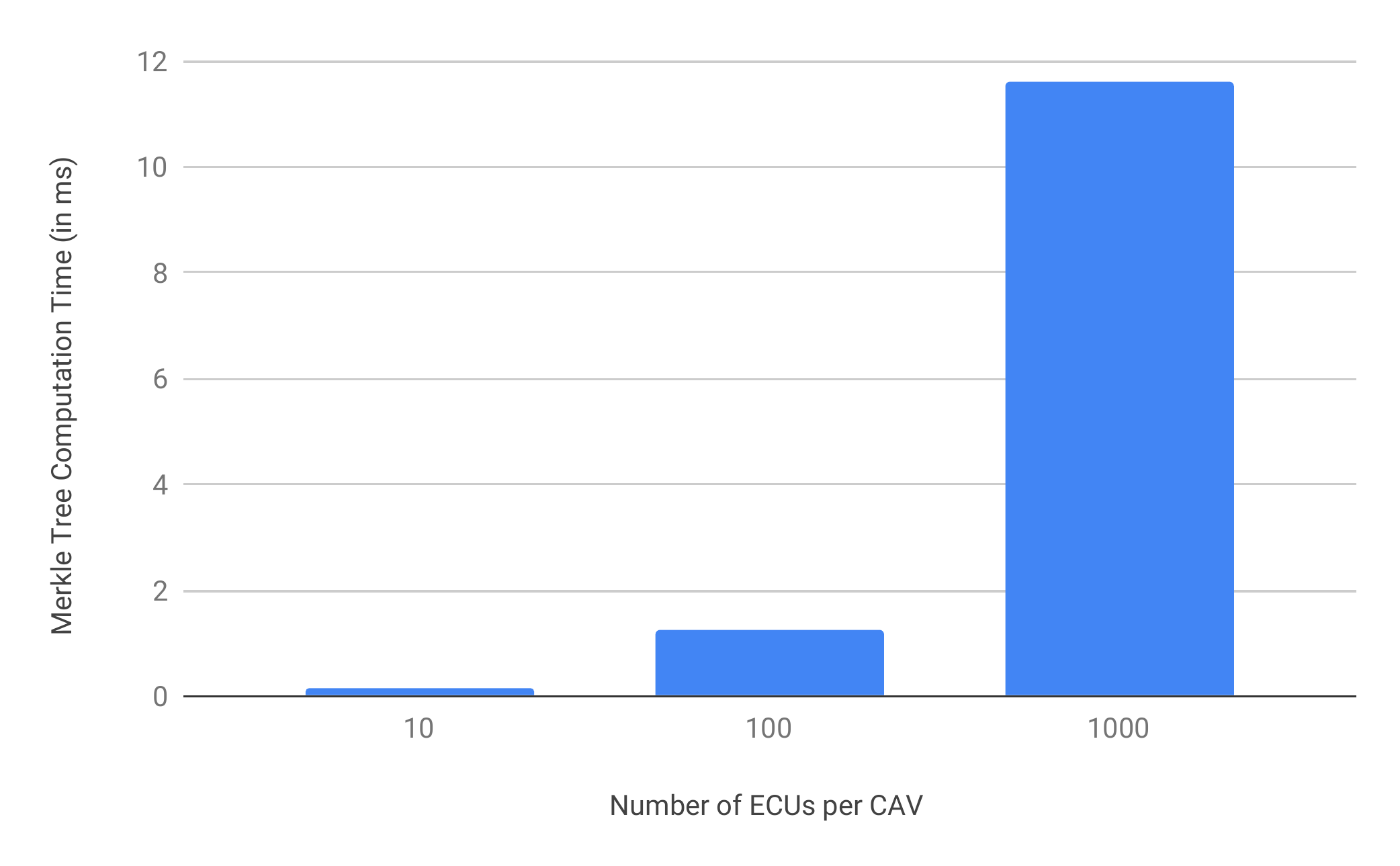}
\caption{Time taken to calculate Merkle tree root}
\label{fig:merkleResult}
\end{figure}
In the next evaluation, we evaluate the time it takes a \textit{CAV} to compute its merkle tree root defined as the cumulative sum of all its ECU hash. According to NXP, a semiconductor supplier for automotive industries~\cite{NXP:2017}, the number of ECUs range from 30 to 100 in a modern vehicle. In this evaluation, we assume that as vehicle functions become more automated, the number of ECUs is likely to increase. Therefore, in our experiments, we vary the number of ECUs from 10 to 1,000.  Figure~\ref{fig:merkleResult} presents the time to compute the Merkle tree root. The results present a linear growth as the number of ECUS increases. In our result, even when the number of ECUs in a \textit{CAV} are 1000,  the time to compute the Merkle tree root is about 12 ms which is still an acceptable time for a highly complex scenario.

\begin{figure}[h]
\centering
\includegraphics[width=0.5\textwidth]{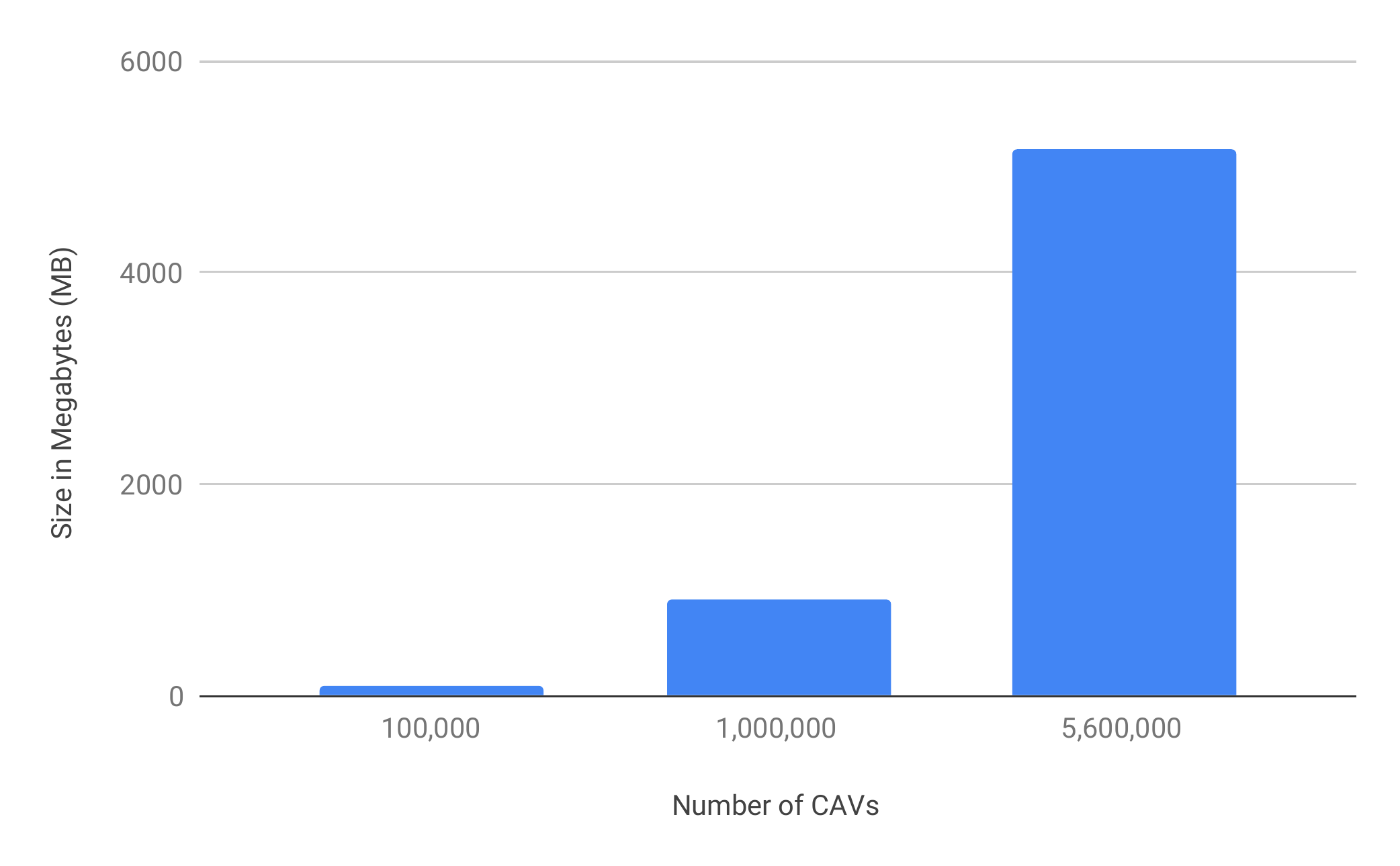}
\caption{Blockchain size}
\label{fig:blocksize}
\end{figure}
In the final evaluation, we consider the amount of storage required by an RSU to store the BC for different number of \textit{CAVs}. To get a realistic picture of required storage, we considered the number of vehicles in New South Wales (NSW), Australia in 2018.  As presented in Figure~\ref{fig:blocksize}, the number of blocks (which represents the number of vehicles)  was changed from 100,000 representing a small city in NSW to 5,600,0000\footnote{5,600,0000 represents the number of cars in the state of New South Wales, according to www.abs.gov.au}.  Based on the results, an RSU must have around 5 Gb to store the BC structure in the state of New South Wales. This result show that it is feasible for an RSU to maintain the BC for all \textit{CAVs} in NSW.

\section{Discussion}
In this section, we provide a further discussion considering the security, Use cases as well a comparative evaluation of B-FERL against related work.

\subsection{Security analysis}
In this section, we discuss how our proposal demonstrates resilience against attacks described in the attack model. \\
\textbf{Fake data:} For this to occur, one or more data generating ECU of a \textit{CAV} would have been compromised. We can detect this attack during the challenge-response activity between the compromised \textit{CAV} and an RSU where the \textit{CAV} is expected to prove the integrity of its ECU by computing its ECU Merkle tree root value. \\
\textbf{Code injection:} Actions executed by service technicians and vehicle manufacturers are stored in the upper tier and could be traced back to them. Vehicle owners are not be able to alter their odometer value as such actions would make the $SS_{ID}$ value different from what is in its record in the lower tier. \\
\textbf{Sybil attack:} The only entities capable of creating entities in the vehicular networks are the verifiers in the upper tier who are assumed to be trusted. A vehicle trying to create multiple entities must be able to create valid blocks for those entities which is infeasible in our approach. \\
\textbf{Masquerade attack (fake vehicles):} A compromised RSU cannot create a block for a \textit{CAV}. As such, this attack is unlikely to be undetected in B-FERL. Also, a \textit{CAV} is considered valid only if its public key exists in the BC managed by the road transport and legal authorities. \\
\textbf{ECU State Reversal Attack: } We address this attack using the random ECU integrity verification challenge. By randomly requesting the values of ECU in a \textit{CAV}, RSUs could detect the reversal attack by comparing the timestamps ECUs against their entries in the lower tier BC. 
Having discussed our defense mechanism, it is noteworthy that while the utilization of a public key introduces a trade-off that compromises privacy and anonymity of a vehicle, the public key is only utilized by a RSU to identify a vehicle in the challenge-response transaction which ascertains the state of a vehicle and does not require the transmission of sensitive and privacy related information. 
\subsection{Use case}
In this section, we discuss the applicability of our proposed solution to the following use cases in the vehicular networks domain: (1) Vehicular forensics, (2) Trust management, and (3) Secure vehicular communication. \\
\textbf{\textit{Vehicular forensics:}} In the liability attribution model proposed for \textit{CAVs} in~\cite{Oham:2018}, liability in the event of an accident could be split amongst entities responsible for the day-to-day operation of the \textit{CAVs} including the vehicle manufacturers, service technicians and vehicle owners. Also, the authors in~\cite{Norton:2017} have identified conditions for the attribution of liability to the aforementioned entities. The consensus is to attribute liability to vehicle manufacturer and technicians for product defect and service failure respectively and to the vehicle owners for negligence. In our proposed work, we keep track of authorized actions of vehicle manufacturers and service technicians in the upper tier and so we are able to identify which entity executed the last action on the vehicle before the accident. Also, with the challenge-response between RSUs and \textit{CAVs} in the lower tier, we are able to obtain historical proof that proves how honest or rogue a vehicle has been in the vehicular network. Consider the \textit{CAV} in Figure 1, if before entering the coverage region of an RSU, an accident occurs, we could generate evidence before the occurrence of the accident in the lower tier that reflects the behavior of the \textit{CAV} and such evidence could be utilized with the accident data captured by the vehicle for facilitating liability decisions. \\
 \textbf{\textit{Trust Management:}} Trust management in vehicular networks either assesses the veracity of data generated by a vehicle or the reputation of a vehicle [19]. This information is used to evaluate trust in the network. However, existing works on trust management for vehicular networks significantly relies on the presence of witness vehicles to make trust based decisions [19-22] and could therefore make wrong trust decisions if there are little or no witnesses available. Also, reliance on witnesses also facilitate tactical attacks like collusion and badmouthing. In our proposal, we rely solely on data generated by a \textit{CAV} and we can confirm the veracity of data generated or communicated by the \textit{CAV} by obtaining such evidence in the lower tier from the historical challenge-response activity between a \textit{CAV} and RSUs as the \textit{CAV} travels. \\
\textbf{\textit{Secure vehicular communication networks:}} Given that the successful execution of a malicious action by a \textit{CAV} reflects that at least one of the \textit{CAV's} ECUs has been compromised and as a result, undermines the security of the vehicular networks. We describe below how our proposal suffices as an apposite security solution for vehicular networks. \\
\textbf{Identifying compromised \textit{CAVs}}: By proving the  state of ECUs in \textit{CAVs}, we can quickly identify cases of ECU tampering and quickly broadcast a notification of malicious presence in the vehicular network to prevent other \textit{CAVs} from communicating with the compromised \textit{CAV}. \\
\textbf{Effective revocation mechanism:} Upon the identification of a malicious \textit{CAV} during the challenge-response activity, Road transport authorities could also efficiently revoke the communication rights of such compromised \textit{CAV} to prevent further compromise such as the propagation of false messages in the network by the compromised \textit{CAV}.
\subsection{Comparative evaluation}
In this section, we comparatively evaluate B-FERL against the works proposed in [9-10], [15], [23] using identified requirements for securing in-vehicle networks.  \\
\textbf{Adversaries}: Identified works are vulnerable to attacks executed by authorized entities (insider attacks) but in B-FERL, we address this challenge by capturing all interactions between all entities responsible for the operation of the \textit{CAV} including the owner, manufacturer and service technician. By recording these actions in the upper tier (BC), we ensure that no entity can repudiate its actions. Furthermore, by proving the state of ECUs in a \textit{CAV}, we are able to identify possible attacks. \\
\textbf{Decentralization:} By storing vehicle related data as well as actions executed by manufacturers and service technicians in the BC, we ensure that no entity can alter or modify any of its actions. Also, by  verifying the internal state of a \textit{CAV} as it moves from one RSU to another, we preserve the security of the vehicular networks. \\ 
\textbf{Privacy:} By restricting access to information to only authorized entities in B-FERL, we preserve the privacy of concerned entities in our proposed framework. \\ 
\textbf{Safety:} By verifying the current state of a \textit{CAV} against its record in the lower tier, we ensure communication occurs only between valid and honest \textit{CAVs} which ultimately translates to secure communications in the vehicular network.

\section{Conclusion}
In this paper, we have presented a Blockchain based Framework for sEcuring smaRt vehicLes (B-FERL). The purpose of B-FERL is to identify when an ECU of a smart vehicle have been compromised by querying the internal state of the vehicle and escalate identified compromise to requisite authorities such as the road transport and legal authority who takes necessary measure to prevent such compromised vehicles from causing harm to the vehicular network. Given this possibility, B-FERL doubles as a detection and reaction mechanism offering adequate security to vehicles and the vehicular network. Also, we demonstrated the practical applicability of B-FERL to critical applications in the vehicular networks domain including trust management, secure vehicular  network and vehicular forensics where we discuss how B-FERL could offer non-repudiable and reliable  evidence to facilitate liability attribution. Furthermore, by qualitatively evaluating the performance of B-FERL, we demonstrate how it addresses key challenges of earlier identified works. Security analysis also confirms B-FERL's resilience to a broad range of attacks perpetuated by adversaries including those executed by supposedly benign internal entities. Simulation results reflect the practical applicability of B-FERL in realistic scenarios. \\
Our current proposal provides security for smart vehicles by identifying when a vehicle becomes compromised and secures the vehicle against possible exploitations by internal adversaries. An interesting future direction would be to consider the privacy implication for a smart vehicle as it travels from one roadside unit to another.

\ifCLASSOPTIONcaptionsoff
  \newpage
\fi

\end{document}